\documentclass[sigconf]{acmart}
\acmConference[HT '20]{HT '20: 31st ACM Conference on Hypertext and Social Media }{July 13--15, 2020}{Orlando, FL}
\acmBooktitle{HT '20: 31st ACM Conference on Hypertext and Social Media,
  July 13--15, 2020, Orlando, FL}
\acmPrice{15.00}
\acmISBN{978-1-4503-XXXX-X/18/06}

\usepackage[]{natbib}
\usepackage{amsmath,amssymb,amsfonts}
\usepackage{algorithmic}
\usepackage{textcomp}
\usepackage{xcolor}
\def\BibTeX{{\rm B\kern-.05em{\sc i\kern-.025em b}\kern-.08em
    T\kern-.1667em\lower.7ex\hbox{E}\kern-.125emX}}
\usepackage[]{url} %Required
\usepackage{graphicx}  %Required
\usepackage{placeins}

\usepackage{colortbl}
\usepackage{nth}
\usepackage{capt-of}
\usepackage{comment}
\usepackage{multirow}
\usepackage{float}
\usepackage{booktabs}
\usepackage{enumitem}
\usepackage{multicol}
\usepackage{graphicx}
\usepackage{subcaption}
\usepackage{cleveref}

\begin{document}
\title{What is BitChute? Characterizing the ``Free Speech'' Alternative to YouTube}

% %\author{Removed for Blind Review}
% \author{Milo Trujillo\textsuperscript{1}, Maur\'{i}cio Gruppi\textsuperscript{1}, Cody Buntain\textsuperscript{2}, and Benjamin D. Horne\textsuperscript{1}\\
%  Rensselaer Polytechnic Institute\textsuperscript{1}, New Jersey Institute of Technology\textsuperscript{2}\\
%  \{trujim,gouvem\}@rpi.edu, cbuntain@njit.edu, horneb@rpi.edu
% }

% \author{Author 1}
% \author{Author 2}
% \email{(author1,author2)@email.com}
% \affiliation{%
%   \institution{Research Institution}
%   \city{City}
%   \state{State}
%   \postcode{Zip}
% }

\author{Milo Trujillo}
\authornote{Both authors contributed equally to this research.}
\author{Maur\'icio Gruppi}
\authornotemark[1]
\email{(trujim,gouvem)@rpi.edu}
\affiliation{%
  \institution{Rensselaer Polytechnic Institute}
  \city{Troy}
  \state{New York}
  \postcode{12180}
}

\author{Cody Buntain}
\email{cbuntain@njit.edu}
\affiliation{%
  \institution{New Jersey Institute of Technology}
  \city{Newark}
  \state{New Jersey}
  \postcode{07102}
}

\author{Benjamin D. Horne}
\email{horneb@rpi.edu}
\affiliation{%
  \institution{Rensselaer Polytechnic Institute}
  \city{Troy}
  \state{New York}
  \postcode{12180}
}

% Bold red text, note to co-authors
\newcommand{\red}[1]{\textbf{\textcolor{red}{#1}}}

\interfootnotelinepenalty=10000 % Make LaTeX try _very_ hard to keep footnotes on one page

\begin{abstract}
In this paper, we characterize the content and discourse on BitChute, a social video-hosting platform. Launched in 2017 as an alternative to YouTube, BitChute joins an ecosystem of alternative, low content moderation platforms, including Gab, Voat, Minds, and 4chan. Uniquely, BitChute is the first of these alternative platforms to focus on video content and is growing in popularity. Our analysis reveals several key characteristics of the platform. We find that only a handful of channels receive any engagement, and almost all of those channels contain conspiracies or hate speech. This high rate of hate speech on the platform as a whole, much of which is anti-Semitic, is particularly concerning. Our results suggest that BitChute has a higher rate of hate speech than Gab but less than 4chan. Lastly, we find that while some BitChute content producers have been banned from other platforms, many maintain profiles on mainstream social media platforms, particularly YouTube. This paper contributes a first look at the content and discourse on BitChute and provides a building block for future research on low content moderation platforms. 
% several potential pathways to BitChute's extreme content from mainstream platforms, including the presence of BitChute content producers on mainstream social media and gaming content. 
\end{abstract}




\maketitle
{\textbf{\small{This paper is supplemental to a short version of the paper published in ACM Conference on Hypertext and Social Media}}}

{\small{\textbf{Warning:} This  paper  contains  disturbing  and  offensive language.  While  we  try  to  discuss  some  of  the  more  disturbing material at a high level, we do not censor language exposed in analysis.}}

\section{Introduction}
In recent years, the online media ecosystem has gained significant attention due to its role in false information spread, radicalizing ideological extremists, and perpetuating malicious hate speech. Due to these rising concerns, several social media platforms, including Twitter, YouTube, and Reddit, have begun efforts to mitigate both false information and hate speech through a variety of methods, including banning users, quarantining communities, and demonetizing content creators. This new approach on platforms that were once more lax in content moderation has led to the proliferation of many alternative social platforms, which harbor banned and demonetized content creators in the name of free speech.

While moving extremists off of large mainstream platforms seems to be a pro-social solution, as it potentially limits exposure to anti-social content, the online cultures created and propagated by fringe platforms can have real-world impacts. For example, in October 2018, a Gab user carried out a mass shooting at a Pittsburgh synagogue after previously announcing the shooting on Gab, where he also participated in anti-Semitic discourse~\cite{mcilroy2019welcome}. Similarly, in August 2019, a mass shooting occurred in El Paso, Texas, where police believe the gunman had previously posted a white nationalist themed manifesto on 8chan~\cite{arango2019nyt}. Unfortunately, these examples only cover a small fraction of recent events in which online hate speech has motivated, incited, or has been connected to offline violence~\cite{hatebase-art}. While it is unclear how causal the role of fringe platforms are in these events, clear connections exist between the online discourse and the stated motivations of these events. 

% On top of these real life impacts, it is still unclear how much of the content produced on low content-moderation platforms stays off of mainstream platforms. It is relatively unexplored what types of information and radicalization pathways exist between the two online worlds. 

These violent events and questions of broader online radicalization have shifted the focus of researchers and media towards these alternative platforms. A selection of this research shows high levels of hate speech and broader impact on the online ecosystem. Specifically, research on Gab has shown that the platform has high levels of hate speech and evidence of alt-right recruiting efforts~\cite{zannettou2018gab}, communities within 4chan, Reddit, and Gab have been shown to spread hateful and racist memes across the web~\cite{zannettou2018origins}, and research on 4chan's ``Politically Incorrect'' board /pol/ has shown evidence of impact on mainstream platforms, including ``raids'' of YouTube comment sections~\cite{hine2017kek,mariconti2019you}. 

Despite this growing attention, one alternative platform, BitChute, has, for the most part, operated under-the-radar, receiving little attention from media or researchers. BitChute is a recently launched video-hosting platform that seeks to provide a ``censorship-free'' alternative to YouTube. BitChute's 2017 launch comes in a wave of other alternative social platform launches: BitChute is an alternative to YouTube, as Gab is an alternative to Twitter, and Voat is an alternative to Reddit. Just as other fringe platforms, BitChute has been said to provide sanctuary for conspiracy theorists, disinformation producers, and hate speech. Specifically, the Southern Poverty Law Center has described BitChute as a ``low-rent YouTube clone that carries an array of hate-fueled material''~\cite{splc1}. While BitChute is currently smaller than communities like 4chan and Gab, recent analysis of Google Trends data suggests the platform is growing in popularity and doing so faster than other alternative platforms (see Figure~\ref{fig:google_trends.pdf}). Furthermore, the immense popularity of social video content points to the impending success of BitChute. According to a 2019 survey, YouTube was used by 73\% of U.S. adults, making it the most widely used social media platform in the United States. In comparison, only 22\% of U.S. adults said they used Twitter, and 11\% of U.S. adults said they used Reddit~\cite{pew1}. This contrast demonstrates the pervasive popularity of social video content, a role which BitChute seeks to fill. As the alternative news ecosystem continues to diverge from mainstream sources~\cite{starbird2017examining} and YouTube commits more to moderation efforts~\cite{youtube2,youtube1}, an alternative space to YouTube will likely be a key platform in this space. If we want to address this increasing polarization in the media environment, an understanding of BitChute's discourse, types of content, and connections with the larger ecosystem are critical to developing such interventions.

To address this knowledge gap and better understand the potential dangers of BitChute, we perform an exploratory, mixed-methods analysis of the video-hosting platform. Our goal with this work is to provide a broad characterization of the platform for future research to build from. To the best of our knowledge, this work is the first characterization of BitChute and its place in the alternative social media ecosystem. Specifically, we address four primary questions: 
\begin{enumerate}[label=\textbf{Q\arabic*.}]
    \item How active are users on the platform and is interest in the platform growing?
    \item What types of videos and content producers does BitChute attract, and what content receives engagement?
    \item What types of discussions happen in BitChute comment threads?
    \item What connections to contemporary social media platforms exist on BitChute?
\end{enumerate}

Our analysis reveals several key findings:
\begin{enumerate}[label=\textbf{R\arabic*.}]
    \item Bitchute is growing in popularity. According to Google Trends, the platform has gained interest faster than other alternative platforms over 2019.
    
    \item \begin{enumerate}[label=\textbf{\alph*.}]
        \item {BitChute is primarily used for news and political commentary, attracting many ``news-like'' channels that provide mostly conspiracy-driven content.}  %There is also a large amount of content from the MGTOW community, an anti-feminist group.
    
        \item Only a handful of channels receive any engagement, but almost all of those channels contain far-right conspiracies or extreme hate speech (i.e. 12\% of the channels receive over 85\% of the engagement).

        \item BitChute contains terroristic, neo-Nazi recruitment and calls to violence, and this content receives engagement.
    \end{enumerate}

    \item Both the videos and comments on BitChute contain high amounts of hate speech, mostly anti-Semitic. Evidence shows the rate of hate speech on BitChute is higher than on Gab, but less than 4chan's ``politically incorrect'' board /pol/.

    \item While some BitChute content producers have been banned from other platforms, many maintain profiles on mainstream social media, particularly YouTube, Twitter, and Facebook.
    
%   \item There are pathways to BitChute through teaser videos on YouTube and BitChute content producers on mainstream social media.

\end{enumerate}

\section{Related Work}
A large body of work exists on alternative or fringe online social platforms and communities. The earliest and largest set of work has focused on the 4chan platform, launched in late 2003. 4chan is an anonymous, ephemeral imageboard platform, known for its ``politically incorrect'' board /pol/ and the various internet memes created on the platform. Work on 4chan has primarily focused on the behavior of /pol/, including studies of hate speech~\cite{hine2017kek}, specific racist discourse~\cite{mittos2019and}, and measurements of its impact outside of 4chan~\cite{hine2017kek,zannettou2017web,zannettou2018origins}. Similarly, researchers have studied fringe Reddit communities, particularly their connections to 4chan. These studies include exploring news shared across Reddit, 4chan, and Twitter~\cite{zannettou2017web}, discourse on Reddit and 4chan during the Boston Marathon Bombings~\cite{potts2013interfaces}, and detecting potential ideological radicalization on the Alt-right Reddit community~\cite{grover2019detecting}. 

More recent work has focused on characterizing the Gab platform, a low-moderation alternative to Twitter. In 2018, two papers characterizing the platform were published~\cite{zannettou2018gab,lima2018inside}. They both found similar conclusions. Namely, Gab is predominately used for political discourse and the users of Gab are strongly conservative leaning. It was also shown that the rate of hate speech on Gab is more than Twitter, but less than 4chan.

Also related is work on YouTube and its potential radicalization pathways. Specifically, Ribeiro et al. provide the first quantitative analysis of user radicalization on YouTube~\cite{ribeiro2019auditing}. Using a dataset that is focused on the Alt-right radicalization pathway, the authors show that users consistently move from milder to more extreme content. There has also been more specific work focused on Jihadist terrorism content and radicalization on YouTube~\cite{kayode2019dataset,conway2008jihadi}, particularly before YouTube's changes in content moderation and recommendation. 

% While YouTube's moderation has significantly improved as of late, it still has issues. Recent work by Papadamou et al. focuses on characterizing and detecting disturbing videos targeting children on YouTube~\cite{papadamou2019disturbed}. The authors show that even with YouTube's current content moderation practices, young children are still likely to encounter disturbing videos through random browsing.

In these studies, the goals have been to characterize extreme behavior on platforms, to find potential pathways to that extreme content, and to understand how specific platform structures that aid extreme content producers. Our work follows a similar set of goals, but on a new, unstudied platform, BitChute. We contribute to the literature by characterizing this unstudied platform, outlining its role in the larger social media ecosystem, and providing a novel dataset for the continued study of the platform. 

\section{Data}
To answer our research questions, we construct a BitChute dataset. Specifically, we have collected a corpus of video metadata and comment data, which we believe includes all videos publicly posted to BitChute between June 28\textsuperscript{th} and December 3\textsuperscript{rd} 2019, outside of brief data outages. We have collected data on 441K videos in two passes: first by acquiring data about each video as it is posted, and then by returning to each video one week later to collect views, comments, and deletion status. The dataset is freely available at: \url{https://dataverse.harvard.edu/dataverse/mela}.

\textbf{Video Data.}
%The front page of BitChute features three tabs: ``Popular", ``Trending", and ``All", where the ``All" category shows all recently uploaded videos on the platform. By polling the ``All" category of the website every five minutes we accumulate a list of URLs representing every uploaded video. We then visit each of these video URLs, gathering information on the video title, description, author, upload date, user-selected category, and user-selected sensitivity score.
BitChute provides a feed of all recently uploaded videos on the platform. By polling this feed every five minutes we accumulate a list of URLs representing every uploaded video. We then visit each video URL, gathering information on the video title, description, author, upload date, user-selected category, and user-selected sensitivity score.

\textbf{Comments, Views, and Deletions.}
One week after each video is posted, we visit the video URL again. At this time we detect whether the video was removed by the user, removed by BitChute, or remains available. If the video has been removed by the user then we receive a 404 response. If the video has been removed by the BitChute administration then we receive an explicit notice that ``This video has been blocked for breaching the site community guidelines, and is currently unavailable." If the video remains available, then we collect the comments and number of views. 

We chose to collect this information after one week as we assumed the majority of the engagement would happen soon after the video is posted. We tested this assumption after data collection by checking the the number views of videos received in the first week and the number of views videos received six months later (on a subset of data from our first week of collection). We found that 56.3\% of views happen during the first week on average. 

For each comment, we store the video URL the comment is associated with, time the message was posted, the message text contents, and the unique user ID of the comment author. BitChute does not host its own comments but instead uses a third-party commenting service called Disqus.\footnote{\url{disqus.com/}} Because the Disqus comments do not necessarily share login information with BitChute, we cannot establish a one-to-one relation between Disqus and BitChute accounts. In total we collected 854K comments from 38K unique commenters.

\textbf{Note on Privacy.} We identify several of the most active content producers on BitChute in this paper. We consider censoring their names to be inappropriate because their patterns of activity on BitChute and across other platforms are necessary for understanding the growth of alternative media ecosystems and the pathway for media spread between platforms. Since users on BitChute post videos for public consumption, often acting as journalists, they should not have an expectation of anonymity.

\textbf{Interruptions in Data Collection.}~\label{sec:cali}
Our metadata collection server was located in California during most of this study, and lost connectivity during the Pacific Gas \& Electric preemptive power shutdowns\footnote{\url{latimes.com/california/story/2019-11-18/another-power-outage-pge-may-shut-down-grid-northern-california}} on October 23\textsuperscript{rd} and November 18\textsuperscript{th}. Additionally, our IP address was blocked from accessing BitChute on October 11\textsuperscript{th} and BitChute itself was down on November 22\textsuperscript{nd}.

\section{BitChute}~\label{sec:bitchute}
In this section, we provide a general description of BitChute and the guidelines on the platform. 

\textbf{What is BitChute?}
BitChute is a peer-to-peer video hosting service founded in January 2017 that claims to provide a service where creators can "express their ideas freely.' As on YouTube, BitChute allows anyone to create an account and upload videos for free, which can be viewed publicly on the website. Unlike YouTube, there is no personalized recommendation algorithm. The extent of recommendation on the platform is a set of ``popular'' videos on the front page and a related video queue when watching a video. This related video queue is selected by the channel owner of the video being watched, hence, it is not algorithmically selected or personalized. %Notably, BitChute hosts video content using WebTorrent \cite{sivek2004webtorrent}, which utilizes client web-browsers to host popular videos, reducing server costs. \footnote{\label{foot:bitchute-guidelines}\url{www.bitchute.com/policy/guidelines/}}'

\textbf{Community Guidelines and Terms of Service.}
BitChute is not entirely without rules and emphasizes certain types of content.
Specifically, their community guidelines forbid child abuse, terrorist material, threats and incitements to violence, and ``malicious use of the platform.'' Users are ``responsible for setting the correct sensitivity of their content.'' BitChute states they will comply with copyright. BitChute's terms of service require users to be at least 16 years old. Note that BitChute's original terms required users to be 13 years of age or older, and the age requirement increased to 16 on September 25\textsuperscript{th}, 2019.

\textbf{Content Moderation}\label{sec:deletions}
BitChute has three sensitivity ratings for videos: ``Normal,'' ``Not Safe For Work (NSFW),'' and ``Not Safe For Life (NSFL).'' The sensitivity ratings are selected by the video uploader. While BitChute claims proper sensitivity ratings will be enforced, it does not seem to be a priority of the platform. Almost all videos are marked with the default rating, ``Normal," with only 35 videos marked ``NSFW," and 19 marked ``NSFL." Interestingly, the description of what should be classified as ``Normal" was changed by BitChute during our data collection. Specifically, a ``Normal" rating was described as ``Content that is suitable for ages 13+'' and on September 25 was changed to ``Content that is suitable for ages 16 and over.'' This change corresponds to age changes in the terms of service, as mentioned above.

Only 51 videos were removed by BitChute during our data collection. All removed videos were full-length movies. We expect that many of these deletions are a result of DMCA requests or BitChute protecting themselves from copyright infringement. 

% \begin{figure}[h]
% \includegraphics[width=0.95\textwidth]{fig/screenshot1.png}
% \centering
% \caption{Screenshot of BitChute community guidelines found on \url{www.bitchute.com/policy/guidelines/}. }
% \end{figure}

% \begin{figure}[h]
% \includegraphics[width=0.95\textwidth]{fig/screenshot2.png}
% \centering
% \caption{Screenshot of BitChute of principles found on \url{www.bitchute.com/policy/guidelines/}. }
% \end{figure}

\textbf{Uploader Options.}
BitChute users can associate a ``YouTube Channel ID" with their BitChute channel. If they add this information, BitChute will automatically clone all the YouTube videos to BitChute as they are posted. This tool significantly lowers the transition barrier from YouTube to BitChute, allowing users to maintain a presence on both platforms at once without regular effort.

\textbf{Monetization.} BitChute makes money primarily through membership and donation. Users can pay for memberships through cryptocurrencies or SubscribeStar. BitChute also hosts sidebar advertisements from Conversant.\footnote{\url{www.conversantmedia.com/}}

%BitChute previously received donation income through Patreon, Indiegogo, and Paypal, but has been banned by all of these platforms \cite{dailydot1}.

Content creators on BitChute can make money through tips and pledges using services like Patreon, PayPal, SubscribeStar, and CoinPayments. Currently, BitChute does not pay content creators through advertisement cuts. 

% Content Creators can give BitChute a list of their accounts on payment platforms including Patreon, PayPal, SubscribeStar, and CoinPayments. If any of this information is provided, then a ``Tip or Pledge" button appears next to the account name when any videos are viewed, allowing viewers to send money directly to the Content Creator without giving a cut to BitChute. Alternatively, BitChute has proposed allowing creators to upload their own sidebar advertisements, and choose whether BitChute's ads, their own ads, or a mixture appear next to their videos. These options represent a distinct form of monetization from YouTube, where users are paid for each ad shown before their videos. Many users post additional information in their video and channel descriptions linking back to donation services or online stores to finance their content outside the options provided by BitChute.

% BitChute is primarily membership and donation supported. Users can pay for a bronze, silver, or gold monthly BitChute membership, which allow users to host multiple channels and create more video playlists.  a subscription platform similar to Patreon or Indiegogo. As of September 2019,  

\begin{figure}[ht!]
    \begin{subfigure}{0.22\textwidth}
        \centering
        \includegraphics[width=\textwidth]{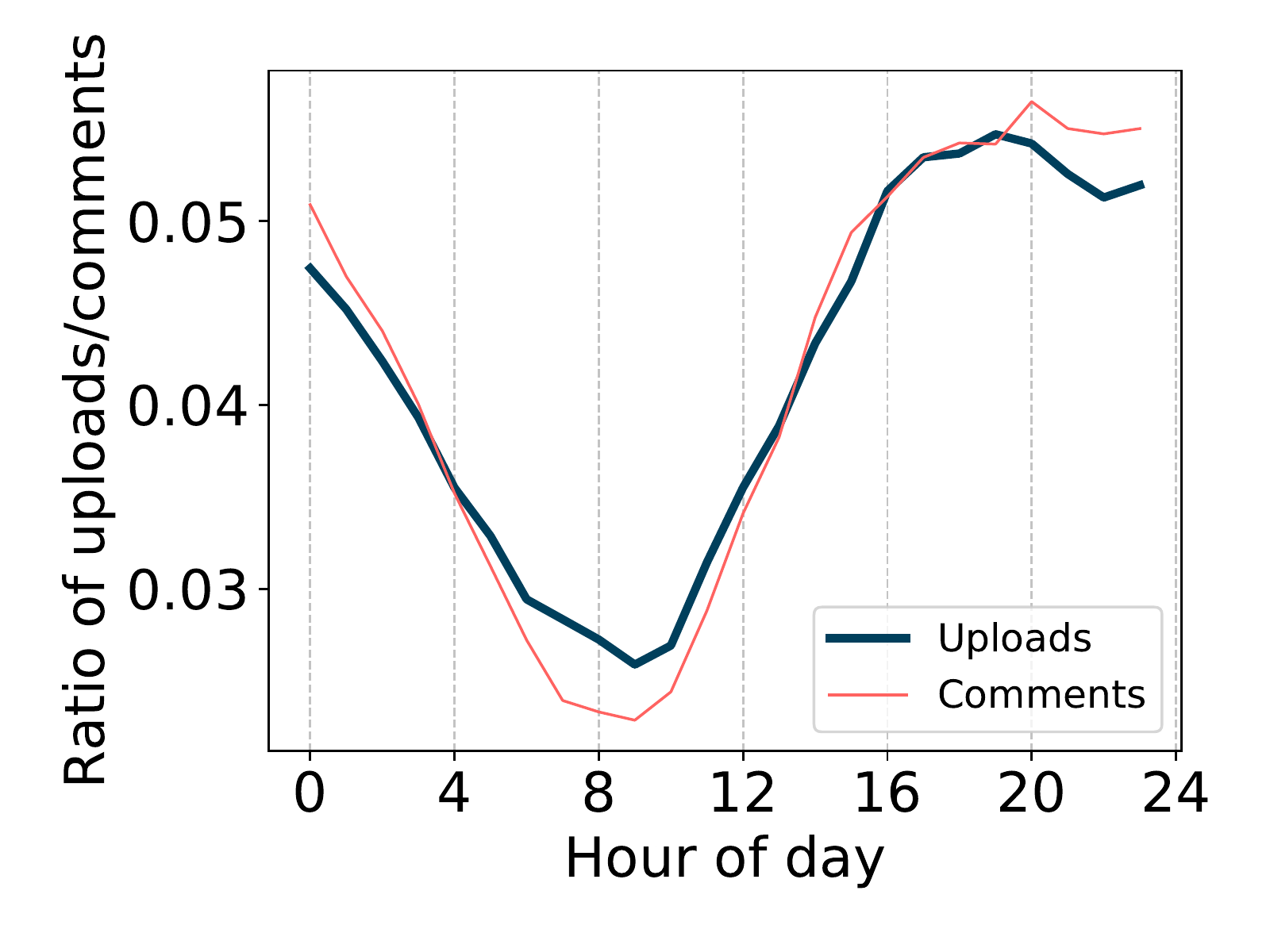}
        \caption{}
        \label{fig:hour.pdf}
    \end{subfigure}
    \begin{subfigure}{0.22\textwidth}
        \centering
        \includegraphics[width=\textwidth]{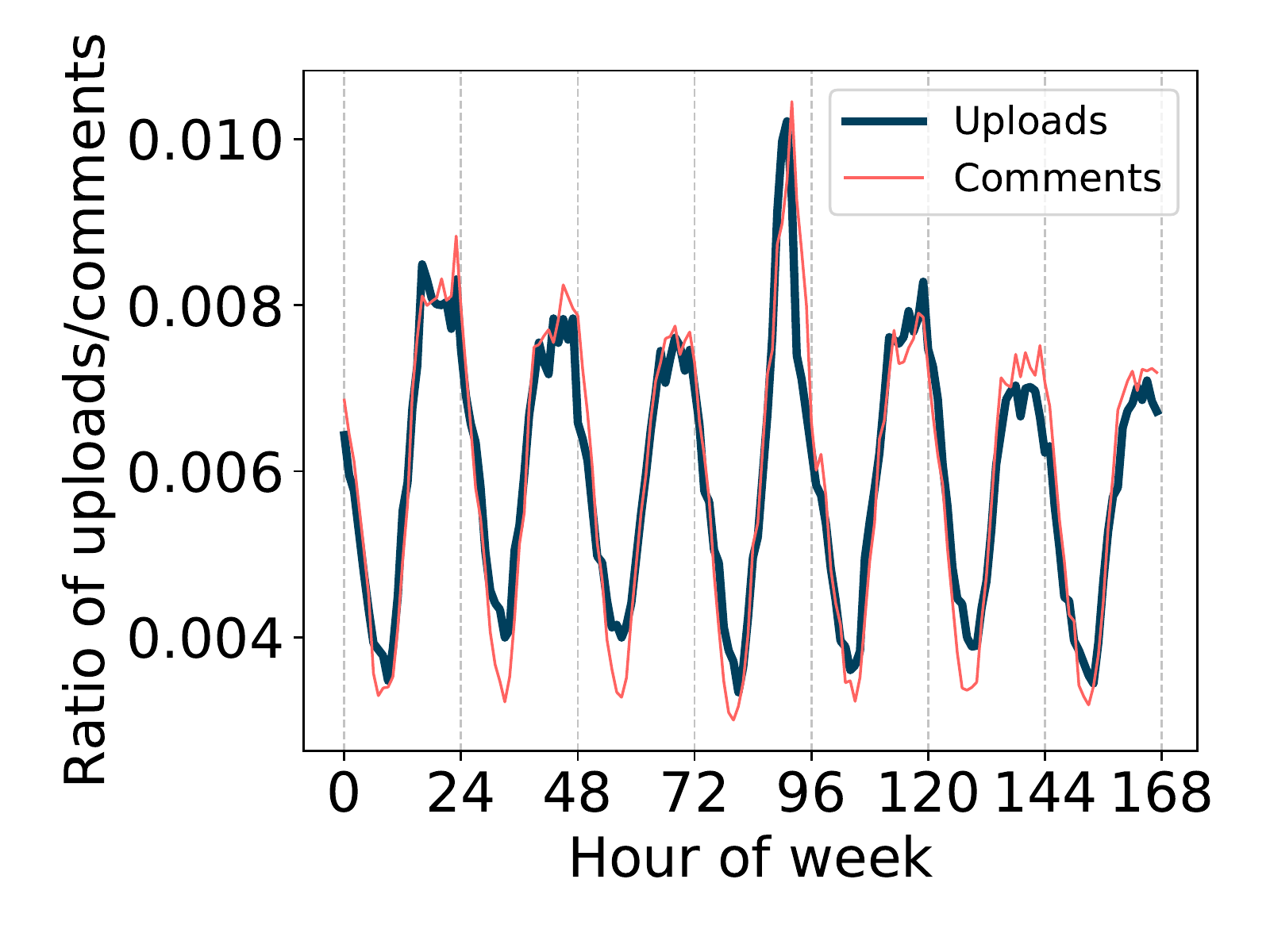}
        \caption{}
        \label{fig:week_hour.pdf}
    \end{subfigure}

    \begin{subfigure}{0.22\textwidth}
        \centering
        \includegraphics[width=\textwidth]{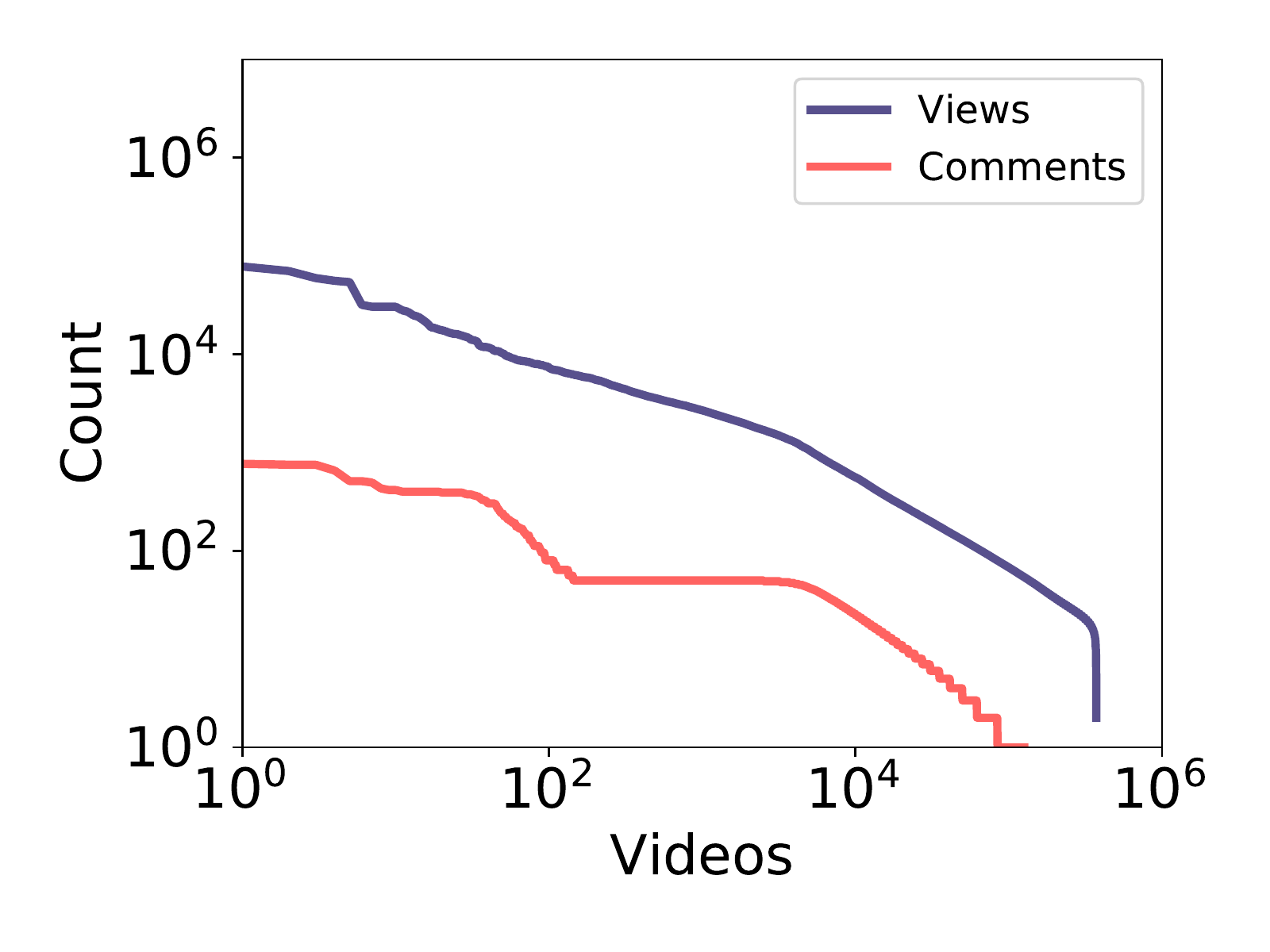}
        \caption{}
        \label{fig:video_dist.pdf}
    \end{subfigure}
    \begin{subfigure}{0.22\textwidth}
        \centering
        \includegraphics[width=\textwidth]{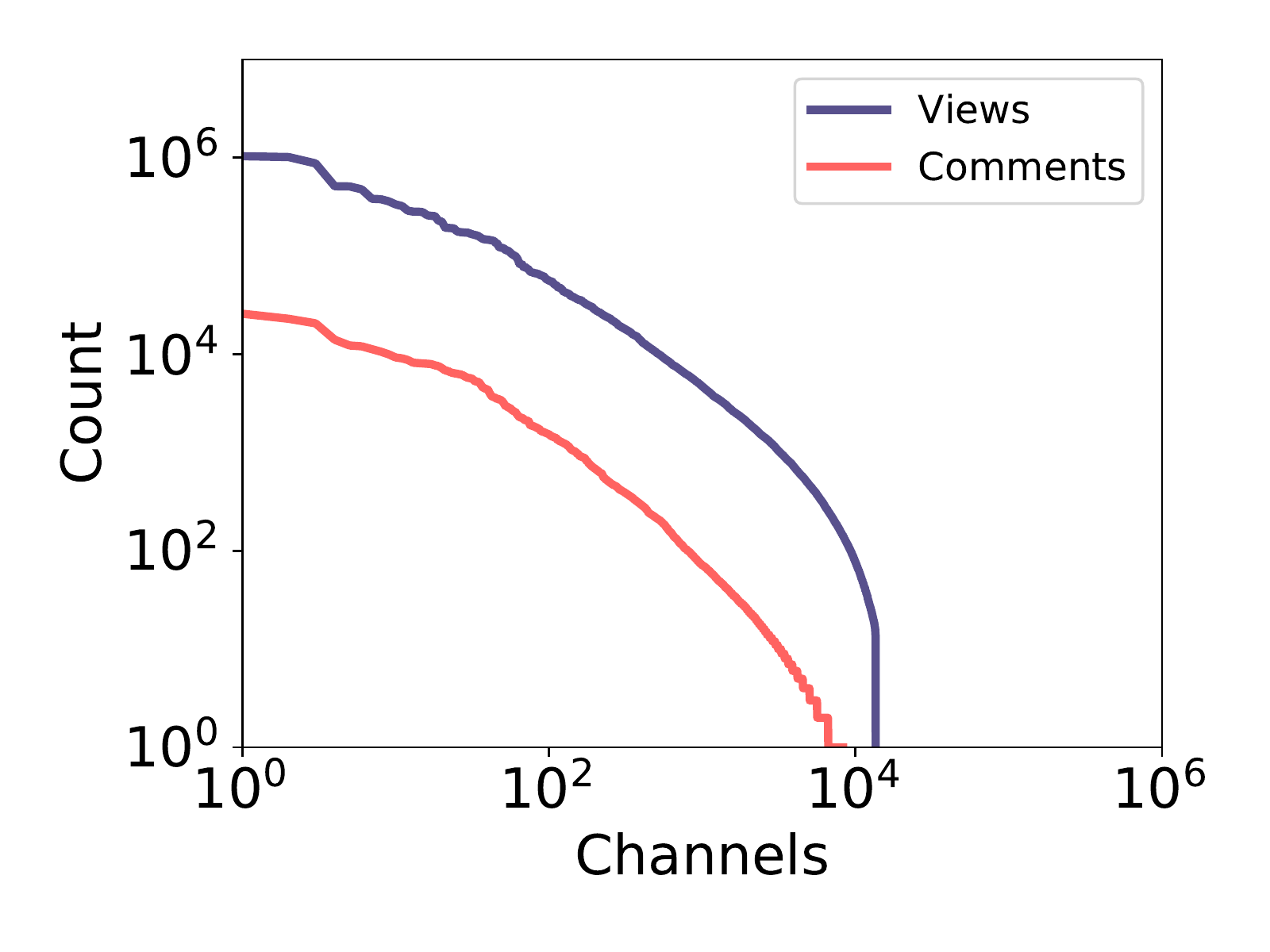}
        \caption{}
        \label{fig:channel_dist.pdf}
    \end{subfigure}
    
    \caption{Top row: distribution of uploads and comments on (a) average per hour of day; (b) average per hour of week where hour 0 is 12 AM on Monday; times are in UTC. Bottom row: (c) observed distributions of video views and comments; (d) Channel views and comments.}
    \label{fig:distribution}
\end{figure}

\section{Activity and Growth}
In this section, we seek to answer \textbf{Q1}: \textit{How active are users on the platform and is interest in the platform growing?} The goal of this section is to provide a basic understanding of the platforms current levels of engagement and its growth over time. 

% and the location of platform users. 

\begin{figure}
    \centering
    \includegraphics[width=0.45\textwidth]{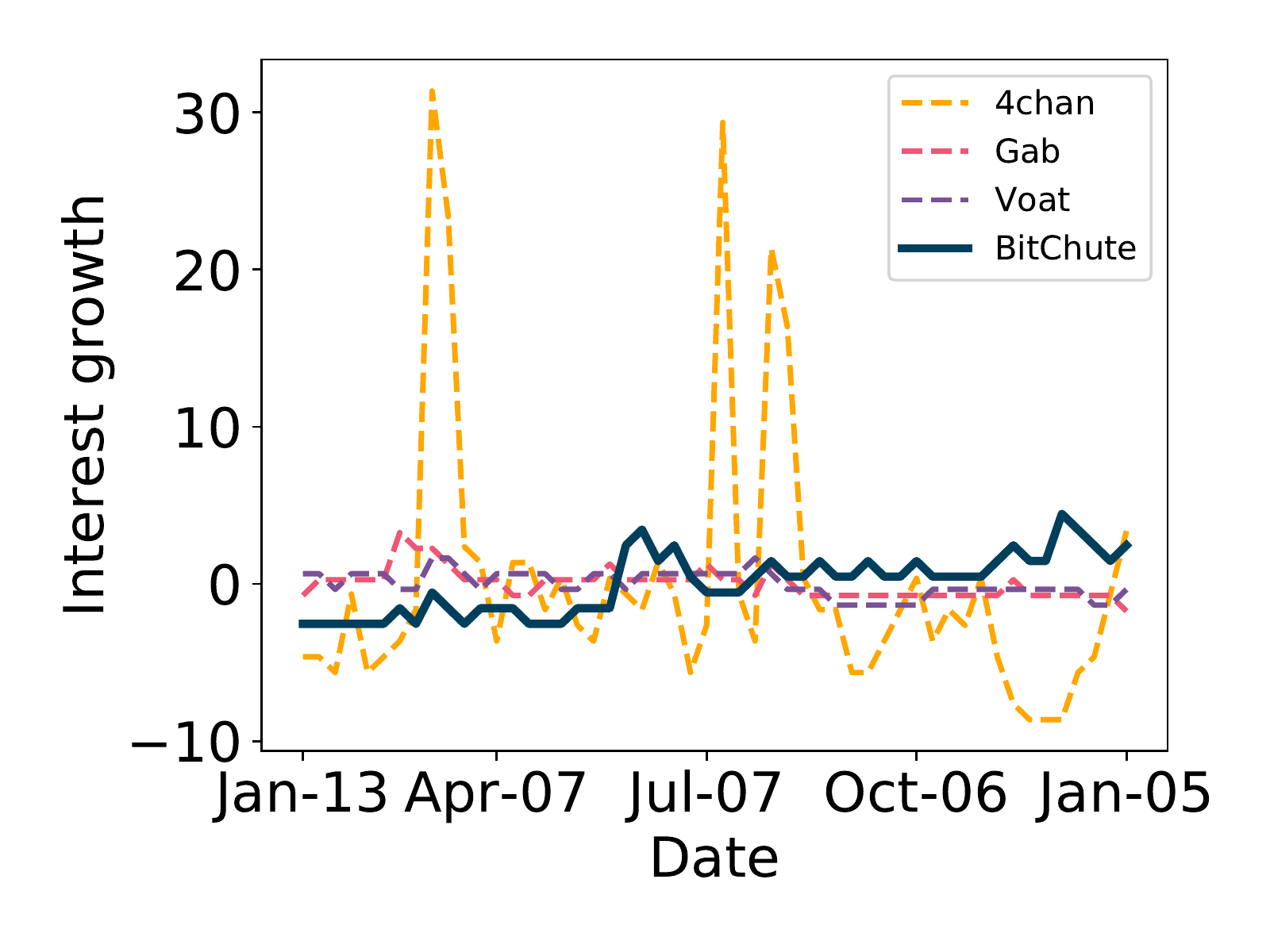}
    \caption{Google Trends interest over time, where the mean is shifted to 0 for comparison.}
    \label{fig:google_trends.pdf}
\end{figure}

\textbf{Activity Distributions.} In Figure \ref{fig:hour.pdf} and \ref{fig:week_hour.pdf}, we show the hours of the day and hours of the week when videos are uploaded to the platform and comments are made on the platform. For the most part, uploading and commenting activity happens during the afternoon and late at night. Both patterns follow those shown on Gab~\cite{zannettou2018gab}.

In Figure~\ref{fig:video_dist.pdf} and \ref{fig:channel_dist.pdf}, we show the distributions of comments per video and comments per channel. As expected, each distribution is roughly a power-law, with a small number of videos receiving a large number of comments. We also find that a small number of users comment very frequently, with the top 5 commenters making 5400 comments and commenting on 2910 unique videos on average.

\textbf{Interest Growth.} We do not see significant upward trends in uploads or views over our dataset. Specifically, the number of videos uploaded per day is roughly consistent through the five months, with around 3K videos uploaded per day. The number of views per day is approximately 300K, with a slight upward trend peaking at just under 400K views over the five months. However, this insignificant growth is likely due to the small time frame in which the data was collected. To gain a better understanding of the platforms growth, we look at Google trends interest over all of 2019, shown in Figure~\ref{fig:google_trends.pdf}. When looking at the Google Trends data, we see that BitChute is growing in interest slightly faster than other alternative platforms like Gab, 4chan, and Voat. This growth primarily happened before our data collection. In fact, the 5 months that our data set covers is roughly flat on the Google Trends interest curve. 

% We also note that all activity on the platform is power-law, with only a handful of users being highly active and many users being minimally active (see Figure~\ref{fig:video_dist.pdf}-\ref{fig:channel_dist.pdf}).

% \textbf{Activity Location.} Most of the platform activity is from the United States. According to Similarweb, 40\% of the traffic is from the United States, 6\% from Germany, and 6\% from Canada.\footnote{\url{www.similarweb.com/website/bitchute.com} accessed 1/3/2020} Alexa rankings similarly shown the United States and Germany as the top countries\footnote{\url{www.alexa.com/siteinfo/bitchute.com} accessed 1/3/2020}. This finding is supported by the language used on the platform. As determined by the \texttt{langdetect} package from \cite{nakatani2010langdetect}, the majority of content is in English (74.89\%).

\section{Content and Engagement}
In this section, we seek to answer \textbf{Q2}: \textit{What types of videos and content producers does BitChute attract, and what content receives engagement?} To answer this, we perform a mixed-methods analysis on data from both the video-level and the channel-level.

\begin{table*}[ht!]
\fontsize{10}{10}\selectfont 
\centering
\begin{tabular}{cccc}
\textbf{Topic \#} & \textbf{Interpretation} &\textbf{Example words} &\textbf{\% Tokens} \\
\midrule
1 & Trump politics & trump, president, impeachment, democrats, biden, clinton, deep state& 18.6\%\\
2 & Mixed Conspiracy Theories & kent, vegas, protesters, hong kong, climate change, media, hitler & 17.6\%\\
3 & Mixed Conspiracy Theories & epstein, jeffrey, truth, dead, body, secret jewish, holocaust, warfare & 17.3\%\\
4 & Alex Jones \& Infowars & infowars, alex, analysis, jones, episode, news, hour, monday & 16.2\%\\
5 & Gaming \& Misogyny & gameplay, mgtow, women, ps4, review, game, vs, team, marvel & 15.6\%\\
6 & News & headline, action, latest, ingraham, news, update, offical, day, night& 14.7\%\\
\bottomrule
&
\end{tabular}
\caption{Topics of video titles as determined by LDA. Word relevance ranking using pyLDAvis was used to interpret topics.}
\label{tbl:topics}
\end{table*}

\subsection{Topics and Categories of Videos} \label{topics}
BitChute provides video uploader-selected categories. As shown in Table~\ref{tbl:cats}, the vast majority of videos are labeled by the uploader as `News \& Politics', with 24.79\% of videos, or `Other', with 33.27\% of the videos. Other is the default category if no category is selected by the uploader. In Table~\ref{tbl:cats}, we also display the engagement in each category, as well as the number of comments that contain hate speech in each category. Details on hate speech in comments can be found in~\ref{sec:hatespeech}.

In order to get a more granular view of video topics on the platform, we analyze the topics found in video titles using LDA~\cite{blei2003latent}. Specifically, we use Scikit-learn's implementation of LDA with $k=6$, where $k$ was chosen using a grid search over model perplexity and the lowest perplexity model was chosen. The model priors are kept as Scikit-learn's default ($1$ divided by the number of components)~\cite{pedregosa2011scikit}. In addition, we use pyLDAvis to interpret topics and topic overlaps\footnote{\url{https://github.com/bmabey/pyLDAvis}}. Results from this analysis are in Table~\ref{tbl:topics}.  

When looking at the LDA topics in Table~\ref{tbl:topics}, we find the majority of content is loosely-related to news and politics (as the user selected categories suggest). There are many words related to Trump politics, news reports, and political conspiracies. There is a significant number of words related to various conspiracy theories. Topic \#2 and \#3 have significant overlap according to pyLDAvis and the only topic clusters to have overlap. In both topics we see words related to hitler and the holocaust, while each topic maintains a distinct set of words related to conspiracy theories. Specifically, in \#2 we see words related to mass shooting events in Las Vegas and Kent State, as well as words relating to Hong Kong protesters. In \#3 we see words related to Jeffrey Epstein's death. Additionally, we find that episodes of Alex Jones radio show, Infowars, are so heavily present on the platform that the show gets its own topic.

Concerningly, we also find a significant amount of gaming content on the platform, providing a potential radicalization pathway~\cite{massanari2017gamergate}. This content is mixed with misogyny, includes key words like: gameplay, marvel, super mario, smash bros, and minecraft, all video games typically targeted to a younger audience. There are also key words related to the anti-feminist, male supremacist group, 'Men Going Their Own Way', including MGTOW and red pill. However, the content receives very little engagement in comparison to news and politics. According to the uploader-selected  categories, `Gaming' only receives 3.87\% of the views and 3.21\% of the comments, while `News \& Politics' receives 48.47\% of the views and 54.04\% of the comments. 

To better understand how well the uploader-selected categories align with the actual video topics, we create additional LDA models for video titles in each category. We find that they align fairly well with the categories themselves (e.g. Gaming has names of video games, News and Politics has names of politicians and political conspiracy theories.). Interestingly, we find that the Education category contains a mix of flat earth videos, Adolf Hitler documentaries, `Men Going Their Own Way' (MGTOW) videos, and videos related to Christianity. The default category `Other' contains a mix of almost every topic, including videos on Adolf Hitler, Alex Jones, Donald Trump, Gaming, and various conspiracies. 

\begin{table*}[ht!]
\fontsize{10}{10}\selectfont 
\centering
\begin{tabular}{ccccc}
\textbf{Category} & \textbf{\% of Videos}  & \textbf{\% of Views}  & \textbf{\% of Comments} & \textbf{\% of Commented Videos w/Hate Comments}\\
\midrule
Other & \textbf{33.27\%} & 27.78\% & 28.58\% & 27.78\%\\
News \& Politics & 24.79\% & \textbf{48.47\%} & \textbf{54.04\%} & \textbf{36.61\%}\\
Gaming & 9.62\% & 3.87\% & 3.21\% &  25.43\%\\
Entertainment & 8.89\% & 6.25\% & 4.37\% & 21.45\%\\
Music & 5.39\% & 2.90\% & 0.94\% &  7.85\%\\
Education & 5.29\% & 4.16\% & 4.37\% & 22.75\%\\
Spirituality \& Faith & 3.10\% & 1.45\% & 1.33\% &  20.59\%\\
Anime \& Animation & 1.76\% & 1.24\% & 0.23\% & 8.00\%\\
Vlogging & 1.58\% & 0.78\% & 0.81\% & 22.83\%\\
Science \& Technology & 1.29\%& 0.72\% & 0.63\% & 10.76\%\\
\bottomrule
&
\end{tabular}
\caption{Video uploader self-selected categories. A video with a hateful comment is defined as any comment under the video having a hate term according to filtered Hatebase lexicon (see Section~\ref{sec:hatespeech}). 'Other' is the default category when uploading a video. Using LDA, we found that the `Other' category contained a mix almost every topic. The following categories have less than 1\% of videos, thus are omitted from the table: People \& Family, Business \& Finance, Sports \& Fitness, Beauty \& Fashion, DIY \& Gardening, Cuisine, Arts \& Literature, Pets \& Wildlife, Auto \& Vehicles, and Travel.}
\label{tbl:cats}
\end{table*}

% \begin{table*}[ht!]
%     \centering
%     \begin{tabular}{cc|cc|cc|cc}
%       \textbf{Domain}&  \textbf{\%}  &  \textbf{Domain} & \textbf{\%} & \textbf{Domain} & \textbf{\%} & \textbf{Domain} & \textbf{\%}\\
%       \hline
%         youtube.com & 25.86\% & nintendo.com & 1.21\% & hungama.com & 0.64\% & wikipedia.org & 0.34\%\\
%         patreon.com & 4.65\% & avantlink.com &  1.18\% & gaana.com & 0.63\% & clkmg.com & 0.33\% \\
%         google.com & 3.20\% & instagram.com & 1.08\% & wynk.in &  0.62\% & reddit.com & 0.33\%\\
%         twitter.com & 2.72\%& mediamonarchy.com & 1.06\% & jiosaavn.com & 0.62\% & twitch.tv & 0.31\%\\
%         therebel.media & 2.33\% & wikimedia.org & 1.03\% & flickr.com & 0.53\% & novinky.cz & 0.27\%\\
%         facebook.com & 2.26\% & amazon.com & 1.00\% & audiolibrary.com & 0.44\% & shopforgamers.com & 0.26\% \\
%         apple.com & 2.13\% & vdare.com & 0.99\% & paypal.com & 0.42\% & endirom.com & 0.26\%\\
%         archive.org & 1.92\% & wordpress.com & 0.94\% & teespring.com & 0.40\% & lisahaven.news & 0.25\%\\
%         rebelnews.com & 1.66\% & bitchute.com & 0.93\% & nichevideogalore.com & 0.34\% & gunstreamer.com & 0.22\%\\
%         blogspot.com & 1.66\% & spotify.com & 0.82\% & soundcloud.com & 0.34\% & newsbud.com & 0.21\%\\
%         \hline
%     \end{tabular}
%     \caption{Top 40 domains used in video descriptions, where $\% =$ number of times domain appeared $/$ total URLs in video descriptions.}
%     \label{tbl:domains}
% \end{table*}

\subsection{Top Viewed Videos}
As discussed, video engagement follows a heavily skewed distribution, with only a handful of videos being viewed very highly. To better understand the most highly engaged with content, we qualitatively describe the top viewed videos in our dataset. Recalling that we collect the number of views a video receives after 1 week of the video being uploaded, videos may receive more views after our collection and analysis. Note, the content in these videos can be disturbing, hence we keep the discussion at a high level. 

\textbf{Soph - Self-proclaimed satirist banned by YouTube.}
The top viewed video in our dataset, with 115K views and 8 comments, is a video entitled ``Pride \& Prejudice'' created by a content creator by the name \texttt{Soph}. This video contains extreme hate speech against homosexuality and Islam. \texttt{Soph} was banned by YouTube for the same video in August 2019. This video has significantly more views in the first week than any other video in our dataset, with the next most viewed video having 37K less views. Due to this video and her other self-proclaimed satire videos, she gained attention from several far-right groups, including being interviewed by Infowars in May 2019 and a spot on the subscription-based video platform freespeech.tv. Since our data collection, all of \texttt{Soph}'s videos have been removed from BitChute and have been moved to freespeech.tv. These videos were removed by \texttt{Soph}, not by BitChute. As of the time of writing, \texttt{Soph} has started to post teaser videos to her full videos behind the freespeech.tv paywall.

% Freespeech.tv is a platform that hosts other video producers who were previously banned from YouTube, such as Milo Yiannopoulos and Proud Boys founder Gavin McInnes (Proud Boys is a group that proclaims to be ``anti-political correctness" and ``anti-white guilt.~\cite{splc5}").

% According to an article by The Daily Dot, \texttt{Soph} got her start on YouTube at 11 years old, previously going by the name \texttt{LtCorbis} or \texttt{LieutenantCorbis}~\cite{dailydot2}. While her commentary has always been ``edgy,'' it originally focused on game-play commentary for games like Call of Duty. Since then her commentary has become much more radical~\cite{buzzfeedsoph}. 

\textbf{Mister Metokur - Hateful commentary against subcultures}
\texttt{Mister Metokur} has two of the top view videos in the top five videos in our dataset, with 78K views, 38 comments and 59K views, 25 comments respectively. One video is hateful commentary against the transgender community and the other is hateful commentary against the furry fandom subculture. Mister Metokur is known as a far-right YouTuber who makes video versions of 4Chan threads. He was previously known as \texttt{Internet Aristocrat} on YouTube, where he  became famous for his videos discussing GamerGate~\cite{massanari2017gamergate}, which have since been removed from YouTube. While he still has a YouTube account, much of his content has been moved to BitChute.

\textbf{Infowars - Conspiracies about September 11th and more}
The third most watched video in our dataset is an episode of Infowars from September 10th, 2019. This video had 70K views and 11 comments. Infowars is long-standing conspiracy theory radio show and fake news website owned by Alex Jones~\cite{splc2}. Infowars and Alex Jones are known for many controversial events, such as false stories on the Sandy Hook shootings for which he was sued for in March 2018. Infowars was removed from YouTube in July 2018. As with most episodes of Infowars, the topics of the video ranged widely. This video discussed multiple conspiracies including a theory that claims September 11th was an inside job by the government and claims about the government turning Americans into robots. 

\textbf{QAnon - Deciphering secret messages from Trump}
The fifth most viewed video in our dataset (with 55K views and and 42 comments) is a video on deciphering secret messages from Trump tweets as a part of the QAnon conspiracy theory. The QAnon conspiracy that started on 4chan, which claims an alleged secret plot by the ``deep state'' against U.S. President Donald Trump~\cite{splc3}. Southern Poverty Law Center has linked a series of violent acts to QAnon supporters and asserts that ``the online community of QAnon supporters is fertile recruiting ground.'' This video is claiming that Trump told his supporters, through secret messages in his tweets, that the ``El-Paso shooting\footnote{\url{wikipedia.org/wiki/2019_El_Paso_shooting}} was a setup by the deep-state.''

\textbf{Terroristic Neo-Nazi recruitment}
While not in the top five viewed videos in our dataset, we find a highly-viewed recruitment video on BitChute from a channel called \texttt{AryanAesthetics}. Specifically, the video is a recruitment video from the Atomwaffen Division, a terroristic, neo-Nazi militia group which has been held responsible for multiple murders and planned violent attacks~\cite{splc4}. The video is 5 minutes long and contains clear calls to Anti-Semitic violence and recruitment. At the end of the video there is an email address to join the group. The video has over 1000 views and 13 comments. The comments contain approving sentiment. 

The channel that uploaded the material does not seem to be directly associated with Atomwaffen Division but contains an array of anti-Semitic content, including Holocaust denial, videos explaining why Hitler was not evil, and videos claiming that the Jews are planning a European genocide. The video uploader claims that they received the recruitment video in a private Telegram channel and re-uploaded it to BitChute but are not part of Atomwaffen Division. %\hspace{200pt} % Add this hspace if the space between the two text columns dissapears

% The channel also hosts combat videos from Azov Regiment, a Ukrainian National Guard regiment infamous for neo-Nazi membership and war crimes.
% Users that comment on \texttt{AryanAesthetics} comment on three other channels, each with over 100 videos and 25000 views, featuring speeches from Adolf Hitler, interviews with serial killers, and a wide range of Anti-Semitic discussion videos. These channels share commenters with additional highly engaged with channels, integrating into the broader BitChute community (see Section~\ref{sec:chnnet} for larger channel relationship analysis).

Upon discovering the video, we submitted it and associated information to the appropriate law enforcement agency.

\begin{table*}[ht!]
    \centering
\begin{tabular}{cc|cc|cc}
\multicolumn{2}{c}{\textbf{Videos}} & \multicolumn{2}{c}{\textbf{Views}} & \multicolumn{2}{c}{\textbf{Comments}}\\
\hline
\textbf{Channel Name} &  \textbf{\#} & \textbf{Channel Name} &  \textbf{\#} & \textbf{Channel Name} &  \textbf{\#}\\
\hline
Scripps & 16330 & styxhexenhammer666 & 1221500 & timpool & 35801\\
voatarchive & 6529 & timpool & 1026769 & infowars & 25877\\
murphycat2012 & 5590 & rongibson & 1000299 & styxhexenhammer666 & 22590\\
drums4reasons & 2724 & infowars & 876543 & highimpactflix & 20472\\
martinbordel & 2297 & voatarchive & 513196 & thequartering & 13638\\
liberum-arbitrium & 2272 & x22report & 507769 & voatarchive & 12293\\
reutersnews & 2105 & turdflingingmonkey & 477060 & nextnewsnetwork & 11936\\
nextnewsnetwork & 1856 & highimpactflix & 380300 & liberum-arbitrium & 11204\\
apologista & 1766 & nextnewsnetwork & 357703 & rebel-media & 10413\\
theresistance1776 & 1629 & democraticrightmovement & 328983 & democraticrightmovement & 9657\\
\hline
\end{tabular}
\caption{Ranking of channels by number of videos, views, and comments. Descriptions of these channels are in Section~\ref{sec:channelrank}.} % add another column with top bigram or phrase?
\label{tbl:channelrank}
\end{table*}

\subsection{Top Viewed and Commented Channels}~\label{sec:channelrank}
So far, we have discussed what types of video content are uploaded to BitChute. In this subsection, we move our analysis to the channel level. Just like on YouTube, each user can create a channel where their uploaded videos are grouped. The majority of the time, these channels have a consistent theme. Users can subscribe to any number of channels, which will provide the user with updates when new videos are uploaded. 

In Table~\ref{tbl:channelrank}, we show the top 10 channels ranked by the number of videos, views, and comments. Almost all of the top viewed channels are self-proclaimed news services, journalist, or political commentary (8 out of 10). These include Tim Pool, an American journalist known for live streaming the Occupy Wall Street protests, Ron Gibson, a regular contributor to Infowars, the Infowars channel itself, and \texttt{Styxhexenhammer666}, a libertarian social commentator known for promoting alt-right talking points. Another news-like channel in the top 10 is \texttt{x22report}, which promotes a wide range of pro-Trump, U.S political conspiracies including QAnon and deep state theories. Overall, the channel is very similar to Alex Jones' Infowars, including advertising for end-of-the-world/doomsday prepping supplies. The channels \texttt{highimpactflix} and \texttt{nextnewsnetwork} fall in a similar vein as \texttt{x22report} but promote less conspiracy theories and more hyper-partisan political commentary. Also in the top viewed channels is \texttt{democraticrightmovement}, a far-right wing group in Ireland, who claim to be ``Anti-EU, Anti-Islam, Anti-Zionism, Anti-Abortion, Anti-Multiculturalism, Anti-SJW, and Pro European Nationalism."

Channels that are not news-like in the top 10 viewed channels are \texttt{voatarchive}, which is an archive of all videos posted to Voat, and \texttt{turdflingingmonkey}, which is a MGTOW channel that promotes many extreme anti-female and misogynistic ideals.

When looking at the channels with the most videos, we see a very different set. These include \texttt{reutersnews}, a channel that copies Reuters news videos from YouTube (this channel does not appear to be associated with Reuters) and \texttt{Scripps}, a channel that uploads local news reports and local weather forecasts. Both of these channels receive almost no engagement. 

% and contain no 'About' information. 

% Another apolitical channel is \texttt{drums4reasons}, which uploads music, mostly focused on classic rock artists.

\begin{figure}[h]
    \centering
    \includegraphics[width=0.45\textwidth]{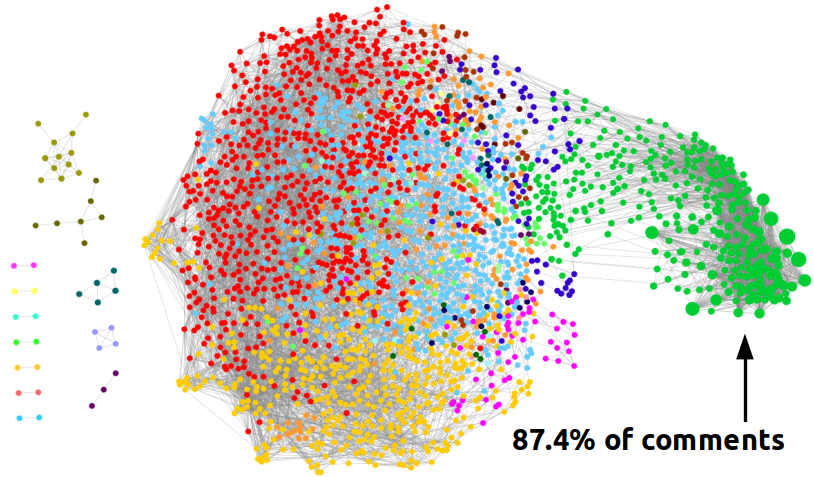}
    \caption {The layout is created using Allegro Edge-Repulsive Clustering in Cytoscape \cite{cytoscape}}
    \label{fig:chn_net}
\end{figure}

\subsection{Channel Audience Overlap}~\label{sec:chnnet}
Next, to understand how popular channels are related, we construct an audience overlap network. In Figure~\ref{fig:chn_net}, we display the network of BitChute channels, where nodes are channels, size of nodes is number of unique commenters, edges are overlapping commenters, and colors are communities. Communities are created using greedy modularity in Python's NetworkX. Edges are weighted by the percentage of shared commenters out of the union of each channel's commenters. We have pruned edges with a weight of under $0.1$, so the network shows links only between channels with at least a 10\% overlap in commenters.

Saliently, the green community only contains 12\% of BitChute channels but receives 88\% of the views and 87\% of the comments from 55\% of the commenters. Over 54\% of those comments contain at least 1 hate speech term. It contains 67\% of the videos on the platform. All of the top viewed and commented channels on the platform are in this community, demonstrating the strong power-law distribution of engagement on the platform. The green community also includes the channels with each of the top 5 viewed videos and the Atomwaffen Division recruitment video. When examining the three highly connected communities (yellow, red, cyan), we see no clear trend of similar content, indicating the divides within the larger subset may be mostly noise. Interestingly, there are several disconnected components with clear themes: one is a Latin MGTOW community, one is a Christian fundamentalist/anti-Islam community, and one is an Irish nationalist community. 

\section{Discussions and Discourse}
In this section, we seek to answer \textbf{Q3}: \textit{What types of discussions happen in BitChute comment threads?} To answer this, we analyze both the hate speech found in comment threads, as well as the topics and top phrases used in comment threads. 

\begin{table*}[ht!]
\fontsize{10}{10}\selectfont 
\centering
\begin{tabular}{cccc}
\textbf{Topic \#} & \textbf{Interpretation} &\textbf{Example words} & \textbf{\% Tokens}\\
\midrule
1 & Governments & government, nation, control, power, rights, freedom, socialism & 14.0\%\\
2 & Policy & money, pay, food, college, study, health, care, vote, work, brexit, don & 13.2\%\\
3 & Video Complements & good, like, thanks, thank you, great stuff, great video, great work & 12.9\%\\
4 & Police \& Protesters & police, gun, right, cops, antifa, law, shot, jail, dead, criminals, rifle & 11.7\%\\
5 & Race & white, jews, black, muslims, racist, fucking, hate, islam, dumb, nigger & 11.1\%\\
6 & Christianity & god, jesus, christ, lord, bible, spirit, love, satan, church, father, sin & 8.5\%\\
7 & Movies \& Media & movie, msm, media, watch, story, star, disney, film, broadcaster & 6.1\%\\
8 & Holocaust & ww2, white, jews, jewish, hitler, zionist, germany, holocaust, nazi & 5.9\%\\
9 & Trump Politics & trump, president, obama, hillary, hoax, clinton, democrats, cia, election & 5.2\%\\
10 & Links to videos \& images & https, com, www, jpg, gif, png, youtube, images, youtu, watch & 5.2\%\\
11 & Locations \& Race & english, irish, england, british, america, uk, south, white, black & 3.6\%\\
12 & Adam Schiff \& Russia & russia, ukraine, oligarch, arms, aerocraft, adam, schiff, collusion & 2.8\%\\
\bottomrule
&
\end{tabular}
\caption{Topics of comments as determined by LDA. Word relevance ranking using pyLDAvis was used to interpret topics.}
\label{tbl:topics_cmts}
\end{table*}

\subsection{Topics in Comments}
To understand what topics are being discussed in comment threads, we again use Scikit-learn's implementation of LDA. This time we generate the model with $k=12$, where $k$ was chosen using a grid search over model perplexity and the lowest perplexity model was chosen. Again, the model priors are kept as Scikit-learn's default~\cite{pedregosa2011scikit}.These topics can be found in Table~\ref{tbl:topics_cmts}. 

The majority of the comments are loosely related to politics. Specifically, we see 27.2\% of the tokens are related to government and policy. We also see comments related to police authority and race relations, which overlap some with the comments on policy. Many of these words relate to a wide variety of alt-right talking points. In addition to these major themes, we see very specific topics in the comments, such as Donald Trump related politics and Adam Schiff related politics. Outside of politics, we also see comments related to Christianity, Movies, and the Holocaust. We also see more generic comments relating to video complements and links to memes. 

\begin{table}[ht!]
\fontsize{10}{10}\selectfont 
    \centering
    
\begin{subtable}[t]{0.45\textwidth}
    \centering
    \begin{tabular}{cc|cc}
    \multicolumn{2}{c}{\textbf{Bigram}}                                      & \multicolumn{2}{c}{\textbf{N-gram ($>$ 2)}}                    \\ \hline
    \textbf{Phrase} & \textbf{\#} & \textbf{Phrase} & \textbf{\#} \\ \hline
    the jew                          & 29302                        & synagogue of satan               & 2557                         \\
    the jews                         & 16306                        & gov doug ducey                   & 1151                         \\
    a jew                            & 10480                        & red ice tv                       & 1102                         \\
    the vid                          & 7839                         & new world order                  & 899                          \\
    the end                          & 6881                         & kingman daily miner              & 872                          \\
    the west                         & 6334                         & red flag law                     & 773                          \\
    the uk                           & 5264                         & ha ha ha                         & 596                          \\
    the eu                           & 4678                         & jew world order                  & 555                          \\
    in america                       & 4506                         & divide and conquer               & 538                          \\
    the democrat                     & 4171                         & world trade center               & 479                          \\ \hline
    \end{tabular}
\end{subtable}

\par\bigskip

\begin{subtable}{0.45\textwidth}
    \centering
    \begin{tabular}{cc|cc}
    \multicolumn{2}{c}{\textbf{Unique Users}} & \multicolumn{2}{c}{\textbf{Unique Channels}} \\ \hline
    \textbf{Phrase}       & \textbf{\#}       & \textbf{Phrase}         & \textbf{\#}         \\ \hline
    the jew               & 3103              & the jew                 & 1657                \\
    the vid               & 2903              & the vid                 & 1635                \\
    the end               & 2609              & the jews                & 1358                \\
    the jews              & 3103              & the end                 & 1274                \\
    the west              & 1782              & on bitchute             & 1258                \\
    the plan              & 1515              & a jew                   & 1105                \\
    the internet          & 1474              & the west                & 895                 \\
    to live               & 1469              & in america              & 826                 \\
    in america            & 1357              & the internet            & 818                 \\
    the uk                & 1350              & the plan                & 814                 \\ \hline
    \end{tabular} 
\end{subtable}

% \begin{tabular}{cc|cc|cc|cc}
% \multicolumn{2}{c}{\textbf{Bigram}} & \multicolumn{2}{c}{\textbf{N-gram ($>$ 2)}} & \multicolumn{2}{c}{\textbf{Unique Users}} & \multicolumn{2}{c}{\textbf{Unique Channels}}\\
% \hline
% \textbf{Phrase} &  \textbf{\#} & \textbf{Phrase} &  \textbf{\#} & \textbf{Phrase} &  \textbf{\#} & \textbf{Phrase} &  \textbf{\#}\\
% \hline
% the jew & 29302 & synagogue of satan & 2557 & the jew & 3103 & the jew & 1657\\
% the jews & 16306 & gov doug ducey & 1151 & the vid & 2903 & the vid & 1635\\
% a jew & 10480 & red ice tv & 1102 & the end & 2609 & the jews & 1358\\
% the vid & 7839 & new world order & 899 & the jews & 3103 & the end & 1274\\
% the end & 6881 & kingman daily miner & 872 & the west & 1782 & on bitchute & 1258\\
% the west & 6334 & red flag law & 773 & the plan & 1515 & a jew & 1105\\
% the uk & 5264 & ha ha ha & 596 & the internet & 1474 & the west & 895\\
% the eu & 4678 & jew world order & 555 & to live & 1469 & in america & 826\\
% in america & 4506 & divide and conquer & 538 & in america & 1357 & the internet & 818\\
% the democrat & 4171 & world trade center & 479 & the uk & 1350 & the plan & 814\\
% \hline
% \end{tabular}
\caption{Top row: phrases used in comments on the platform, ranked by frequency. Bottom row: number of unique commenters who have used the phrase, and number of unique channels the phrase appears in. All phrases are extracted using AutoPhrase and a quality score greater than $0.80$.} 
\label{tbl:commentphrases}
\end{table}

\subsection{Hate Speech in Comments}~\label{sec:hatespeech}
Next, using the Hatebase dictionary~\cite{hatebase}, we assess the amount of hate terms used in the comments on BitChute videos. We found that 75.09\% of comments contained at least 1 hate speech term and the average number of hate terms per hate comment is 5.93. When breaking these terms down by the type of hate, we find that over 45\% of the hate terms used are ethnicity-based hate, followed by 23\% of hate terms used being gender-based hate, and 13\% of hate terms used being class-based hate. Of all videos with or without comments, 21.44\% of videos contain hate speech in the comment thread. Hate terms that frequently occur include `ape' (27K times), `kike' (22K times), `(((' (14K times), `tard' (14K times), and `nigger' (10K times). The triple parenthesis `(((' is a common dog-whistle or replacement word used when talking about Jews. Typically it is used as (((they))) or (((them)))~\cite{finkelstein2018quantitative}.

As pointed out in previous literature, one problem with Hatebase is that it contains many terms that are only hateful in specific contexts, which may inflate the hate speech measurement. To combat this, previous work \cite{hine2017kek} has created a filtered version of the Hatebase lexicon. The filtered lexicon was created by ``manually examining the [Hatebase] list and removing a few of the words that were clearly ambiguous or extremely context sensitive''~\cite{hine2017kek}. This filtering process reduces the size of the Hatebase dictionary from 1526 terms to 1027 terms. 

To compare the numbers reported in previous literature, we acquired this filtered dictionary directly from the authors of \cite{hine2017kek} and recompute our hate speech statistics. When recomputing the proportion of hate terms used in comments, we found that 10.03\% of comments contained at least 1 hate speech term and the average number of hate terms per hate comment is 1.53. While this is a significant decrease in hate speech compared to the numbers from the original lexicon, it is still a significant rate of hate speech. Specifically, in comparison to the numbers reported in \cite{zannettou2018gab} and \cite{hine2017kek}, BitChute comments contain 4.44 times the rate of hate terms as in Twitter posts and 1.85 times the rate of hate terms seen in Gab posts, but the rate of hate terms is 1.2 times less than the rate on 4chan's politically incorrect board, /pol/. 

In Table \ref{tbl:cats}, we show the percent of commented videos that contain hate speech per category using this filtered lexicon.

% However, it is important to note the platforms are different in their medium and current scale, which could account for some of these differences (i.e. the primary medium on BitChute is video, but we measure comments, while Gab is mostly text).

% Additionally, the platforms are different in their medium and scale, which could account for these differences (i.e. the primary medium on BitChute is video, but we measure comments, while Gab is mostly text). In Table~\ref{tbl:cats}, we show the number of commented-on videos with hate speech for each topic category.

\begin{table*}[ht!]
    \centering
    \begin{tabular}{c|c|c}
        \textbf{Channel Name} &  \textbf{Other Social Media Platforms} & \textbf{Banned}\\
        \hline
        styxhexenhammer666 & YouTube, Twitter, Gab, Facebook, Minds, Dailymotion, Brighteon & no\\
        timpool & YouTube, Twitter, Gab, Instagram, Minds, Facebook & no\\
        rongibson & Facebook, Gab, Brighteon, Steemit & yes\\ 
        infowars & Periscope, Twitch, Gab, Minds, Oneway, Scoop.it, ello.co & yes\\
        x22report & YouTube, Twitter, Gab, Facebook, Reddit, Steemit & no\\
        turdflingingmonkey & dlive.tv, YouTube, Discord, Instagram, altCensored & yes\textsuperscript{\textbf{\dag}}\\
        highimpactflix & YouTube, Twitter, Facebook, Instagram, Gab, Minds, Steemit & no\\
        nextnewsnetwork & YouTube, Facebook & yes\\
        democraticrightmovement & Twitter, Gab &  no\\ 
        \hline
        \addlinespace[1ex]
    \multicolumn{2}{l}{\textsuperscript{\textbf{\dag}}\small{Channel has been banned by YouTube, but has an alternative account on YouTube.}}
    \end{tabular}
    \caption{What other social media platforms each of the top 10 viewed channels are present on as of December 31st 2019. We also include in the channel has been banned from any other social media platforms, particularly Twitter or YouTube. We do not include the channel voatarchive as it has no other social media accounts.}
    \label{tab:topchannels_socialmedia}
\end{table*}

\begin{table}[ht!]
\fontsize{10}{10}\selectfont 
    \centering
    \begin{tabular}{cc|cc}
      \textbf{Domain}&  \textbf{\%}  &  \textbf{Domain} & \textbf{\%}\\
      \hline
        youtube.com & 25.86\% & nintendo.com & 1.21\% \\
        patreon.com & 4.65\% & avantlink.com &  1.18\% \\
        google.com & 3.20\% & instagram.com & 1.08\% \\
        twitter.com & 2.72\%& mediamonarchy.com & 1.06\% \\
        therebel.media & 2.33\% & wikimedia.org & 1.03\% \\
        facebook.com & 2.26\% & amazon.com & 1.00\% \\
        apple.com & 2.13\% & vdare.com & 0.99\% \\
        archive.org & 1.92\% & wordpress.com & 0.94\% \\
        rebelnews.com & 1.66\% & bitchute.com & 0.93\% \\
        blogspot.com & 1.66\% & spotify.com & 0.82\% \\
        \hline
    \end{tabular}
    \caption{Top 20 domains used in video descriptions, where $\% =$ number of times domain appeared $/$ total URLs.}
    \label{tbl:domains}
\end{table}

\subsection{Most Frequent Phrases in Comments}
Next, the support both the topic findings and the hate speech analysis, we extract phrases from the full comment corpus using AutoPhrase~\cite{shang2018automated}, a positive-only distant training method for phrase mining. We use a standard set of Wikipedia quality phrases for model training. With high quality phrases extracted, we compute the frequency of each phrase with a quality score above $0.80$. We display the top most frequent bigram phrases, top most frequent greater than two word phrases, and the phrases used by the most unique commenters in Table~\ref{tbl:commentphrases}.  

When looking at Table~\ref{tbl:commentphrases}, it is clear that the platform contains a large amount of anti-Semitic hate speech and conspiracy theories, with 4 of the comments in the top 15 most frequent bigrams and 2 of the comments used by the most unique commenters referring to anti-Semitic language or concepts. Furthermore, when diving deeper into the meaning of the more than two word phrases, we find many anti-Semitic phrases. For example, ``synagogue of satan'' is referring to Jewish people being of satan and is used in conspiracies about Jewish controlled media companies (i.e comments calling companies ``synagogue of satan'' or ``SOS'' companies). Another example is ``new world order,'' which is a call to action that is directly opposed to the phrase ``jew world order.'' Qualitatively, we also see anti-Semitic memes in the same comment threads, such as variants of the ``Happy Merchant'' meme~\cite{finkelstein2018quantitative}. 

We do find that some of the top greater than two word phrases are ``copypasta,'' as one or more users copy and paste a block of text that contains the phrases often (in particular ``gov doug ducey''). However, when looking at each users full comment activity, it seems they are not bots, as they participate in conversations outside of the copypasta. 

\section{Connections to the Outside}
In this section, we seek to answer \textbf{Q4}: \textit{What connections to contemporary social media platforms exist on BitChute?} To answer this, examine the external-facing links in video descriptions and the profiles of the top channels on BitChute.

\subsection{Links to Other Domains in Video Descriptions}
In Table~\ref{tbl:domains}, we show the top 20 domains found in video descriptions after expanding all shortened URLs. We found that 52.29\% of video descriptions contain at least one URL, and video descriptions with a URL contain 3.53 URLs on average. In the top 40 domains found, 6 are other social platforms, 4 are related to monetary support, and 13 are news-like websites or audio platforms to hear news/commentary content. YouTube is the most frequently used domain by a large margin, with 25.86\% of links being links to YouTube. Many channels link back to their former (and sometimes demonetized) or ongoing YouTube channels. Some channels maintain extreme content on BitChute and less extreme content on YouTube. 

Despite the vast majority of links on BitChute pointing to YouTube, YouTube is not the top traffic driver of BitChute. According to both Similarweb and Alexa, top referrals to BitChute include Infowars, Gab, Voat, The Gateway Pundit, and ZeroHedge. Saliently, according to Alexa, Infowars drives magnitudes more traffic to BitChute than the rest. Infowars provides 14.4\% of BitChute traffic while Gab provides 4.2\% and Voat provides 2.4\%.

\subsection{Content Creators' Presence on Other Social Platforms}
Another way to assess links from BitChute to other social media platforms is to match channel profiles across platforms. Specifically, we manually find other social media accounts for the top 10 viewed channels on BitChute and checked if the channel has been previously banned by another social media platform. These results can be found in Table~\ref{tab:topchannels_socialmedia}.

All prominent accounts have a presence on other social media platforms. Many accounts mirror their YouTube content to BitChute, giving them a presence on both a platform with a wider audience and a platform with viewers more aligned with their content. About half of the top accounts have been banned from other social media platforms, pushing them to more fringe communities like BitChute. We also note there are many other small video-hosting sites that these channels are on, but they lack the social structure that platforms like YouTube and BitChute have.

\section{Conclusion and Discussion}
In this paper, we provide the first large-scale characterization of the video hosting platform BitChute. We analyze 441K videos, 15K channels, and 854K comments from 38K unique commenters over five months in 2019. Our analysis reveals that the vast majority of videos and channels are news-like, commentating on current events, politics, and a wide range of conspiracies. The platform's strong focus on politics makes it similar to Gab~\cite{zannettou2018gab}. In fact, many of the content producers on BitChute also have accounts on Gab. We found that the platform's engagement and activity is highly skewed, with only 12\% of channels receiving 88\% of the views and 87\% of the comments on the platform. The channels in this 12\% are heavily focused on news-like commentary and conspiracies.

We found that the content in both the videos and comment threads contains a high amount of hate speech. Our LDA topic analysis and qualitative analysis of videos revealed many anti-Semitic conspiracy theories and targeted hateful commentary towards women, minorities, and subcultures. Our text-based analysis of the comments on the platform also aligned with this hate-fueled theme, showing that 75\% of comments in our dataset contained at least one hate speech term and 21\% of videos contained hate speech in the attached comment thread. When analyzing frequent phrases used in comments, we again found high levels of hate speech, mostly anti-Semitic. Furthermore, we found a recruitment video for the terroristic, neo-Nazi militia, Atomwaffen Division, on the platform. This video was surrounded by other anti-Semitic content and calls to violence. 

Our analysis also reveals several connections between mainstream platforms and BitChute. Many of the top content producers on the platform maintain a social media presence outside of BitChute, mostly on YouTube and Gab. We found that over 25\% of URLs found in video descriptions point back to YouTube. Qualitatively, we found that several of these linked YouTube channels are less extreme versions of the content hosted on BitChute or the content producers maintain disjoint sets of videos on each platform. We found that only one of the five top channels that were not banned from YouTube elected not to maintain a YouTube channel.
% , and one banned channel maintains an alternative channel on YouTube to evade their ban.

In addition, we found a large amount of gaming content on the platform and content producers involved in the 2014 Gamergate controversy. It has been suggested that gaming communities can be pathways to toxicity and are targeted for assimilation into alt-right social movements~\cite{massanari2017gamergate,lindvall2018they}. Having gaming communities on such an aggressive platform like BitChute may create radicalization pathways, though their lack of engagement in our dataset suggests this is not presently an active avenue for indoctrination. However, more in-depth analysis of these potential pathways is needed to understand their effects. While this work provides some analysis of connections to and from the platform, it does not directly address user movement to the platform. This is left up for future work.

Further work should be done to understand BitChute's placement in the larger alternate-media ecosystem and its effects on the web as a whole. For example, it is still unknown and difficult to assess user movement from YouTube to BitChute, particularly for those participating in comment threads. In general, it is also unknown if these information consumers were already participating in aggressive hate speech or if they were assimilated to these communities. Similarly, this work did not directly assess the role of alternative news sources on BitChute~\cite{starbird2017examining}. Using external sources, we found that three alternative news sites drive traffic to BitChute, suggesting that alternative news may play a major role in BitChute video consumption.

% There is also room for work on the impacts of deplatforming. It is still not well understood how effective moving extreme content producers off of major platforms is. On one hand, there is less extreme content on YouTube due to deplatforming. On the other hand, platforms like BitChute have been created, which may create more extreme echo chambers and may lead to offline violence. These concerns merit further investigation; specifically, work in understanding the long-term impacts of mitigation techniques, such as banning users versus quarantining communities. 

In conclusion, BitChute appears to be an increasingly popular video-sharing platform, especially for the dissemination of news and political content. As is common among fringe and alternative platforms, much of BitChute's content is politically extreme and hateful, even more so than one might find on similar alternative platforms like 4chan or Gab. Concerningly, BitChute contains a more diverse mix of content than Gab, including much more gaming and entertainment content, increasing the likelihood that an unradicalized individual may be incidentally exposed to more extreme views. This exposure is especially troubling given the large volume of hate speech, white nationalist recruitment, and calls for violence present on BitChute. 

Our hope with this work is to create a building block to study activity on this platform, to study this platforms interaction with alternative news producers, and to study this platforms interaction with other low content moderation platforms like Gab and 4chan. 

% There is also interest in the financial assets of BitChute. At the time of writing, BitChute self-reports a monthly income of \$13,000, of which we can account for \$11,200 via their SubscribeStar membership subscriptions, and assume the remainder comes from cryptocurrency donations and advertisement income. It remains unclear what BitChute's operational costs are, or who the major financial contributors to the platform are. Of particular interest are any financial ties between BitChute and other platforms hosting significant hate-speech, which could indicate broader coordination. There are also related questions of if and how long-term, potentially state-sponsored, disinformation campaigns are utilizing platforms like BitChute. 

\bibliographystyle{ACM-Reference-Format}
\bibliography{references}
\end{document}